\documentclass[epj]{svjour}
\usepackage{graphics}
\begin{document}
\title{Experimental evidence for the interplay between individual wealth and transaction network}
\author{Jie-Jun Tseng\inst{1,}\thanks{email: gen@phys.sinica.edu.tw} \and Sai-Ping Li\inst{1} \and Sun-Chong Wang\inst{2}
}                     
\authorrunning{J.J. Tseng et al.}
%
%
\institute{Institute of Physics, Academia Sinica, Taipei 115 Taiwan \and
  Institute of Systems Biology and Bioinformatics,
  National Central University, Chungli 320 Taiwan}
\date{Received: date / Revised version: date}
%
\abstract{
We conduct a market experiment with human agents
in order to explore the structure of transaction networks and
to study the dynamics of wealth accumulation.
The experiment is carried out on our platform for 97 days
with 2,095 effective participants and 16,936 times of transactions.
From these data,
the hybrid distribution (log-normal bulk and power-law tail) in the wealth is observed
and we demonstrate that the transaction networks in our market are always scale-free and disassortative
even for those with the size of the order of few hundred.
We further discover that the individual wealth is correlated with its degree by a power-law function
which allows us to relate the exponent of the transaction network degree distribution
to the Pareto index in wealth distribution.
\PACS{
      {87.23.Ge}{Dynamics of social systems} \and
      {89.65.Gh}{Economics; econophysics, financial markets, business and management} \and
      {89.75.Da}{Systems obeying scaling laws} \and
      {89.75.-k}{Complex systems}
     } 
} 
\maketitle
\section{Introduction}
It has been widely observed that the distribution of wealth
among individuals in various economies follows a remarkably simple pattern, namely,
a power-law tail for the rich (also known as Pareto's law) and
a log-normal distribution for the rest~\cite{Pareto,Gibrat,INCOME,WDBOOK}.
In other words, the wealth distribution $P_w(w)$ can be expressed as
\begin{equation}
  P_w(w)  \sim \Big\{
    \begin{array}{ll}
      w^{-(\alpha+1)} &~~\hbox{for $w\ge w_{*}$},\\
      \exp [-\frac{(\log w - \mu)^2}{2 \sigma^2}] &~~\hbox{for $w< w_{*}$},
    \end{array}
\end{equation}
where $w_{*}$ denotes the threshold wealth for the transition
between the log-normal and power-law distribution.
$\alpha$ is a time-dependent parameter called {\it Pareto index}
ranging from 1 to 2 for different countries,
while $\mu$ and $\sigma$ denote the mean and standard deviation of $\log w$, respectively.

In order to explain the empirical distribution,
several attempts have been made to model the dynamics of wealth accumulation process.
For instance, some models describe the process as an multiplicative stochastic process (MSP)
where the agents can interact with each other through a transaction network.
By further introducing the additive noise and boundary constraints,
it has been shown that these models can generate the power-law tail
in the wealth distribution~\cite{Bouchaud,Solomon}.
The kinetic exchange model is another example,
which maps a closed trading market to a gas-like many-body system~\cite{Yakovenko}.
This type of model is capable of producing the wealth distribution resembling
the pattern observed in reality, and has been well studied recently~\cite{Chatterjee,Caon,Chatterjee2}.

In MSP models, an underlying transaction network is introduced
to describe the interactions among agents and the effects of the network topology
on the wealth accumulation have also been explored.
Until recently, several network topologies have been tested,
including small-world networks~\cite{NW1}, scale-free networks~\cite{NW2,NW3},
heterogeneously linked networks~\cite{NW4} and family networks~\cite{NW5}.
It has been shown that either the log-normal or the power-law distribution arises
when the network is homogeneous,
therefore a heterogeneously connected network is needed
to generate the hybrid distribution~\cite{NW7}.
However, since it is difficult to study these networks empirically,
the further knowledge regarding them is still lacking.
An alternative approach is proposed by Diego Garlaschelli and his colleagues,
where they study the interplay between the World Trade Web (WTW)
and the gross domestic product (GDP) of world countries~\cite{WTW1,WTW2,WTW3}.
They have found that the WTW possesses nontrivial topological properties
and is tightly related to the GDP of world countries.
However, the size of the network is bounded by the number of countries in the world,
therefore some characteristics of the network might not be observed due to this limitation.
In order to have a better understanding about
the interplay between transaction networks and wealth accumulation process,
we here propose yet another approach,
namely by analyzing the data from the market experiment on our platform.

In what follows, we first introduce our platform and describe how the market experiment is performed.
We then report the findings for the topology of the transaction networks
and demonstrate the observed relationship between
the network topology and the individual wealth in our experiment.
Finally, we summarize all the findings and propose a possible implication
in explaining the wealth accumulation process.

\section{Experiment}
We build a Web-based platform - Taiwan Political Exchange (TAIPEX) which
can perform the market experiment with anonymous volunteers from the Web.
Including the latest experiment ending on March 2008,
there have been five runs of official experiments performed on our platform~\cite{WI,QF,NT}.
The architecture of TAIPEX and the details about the market experiment are explained as follows.
\subsection{TAIPEX platform}
\begin{figure}
  \resizebox{0.46\textwidth}{!}{\includegraphics{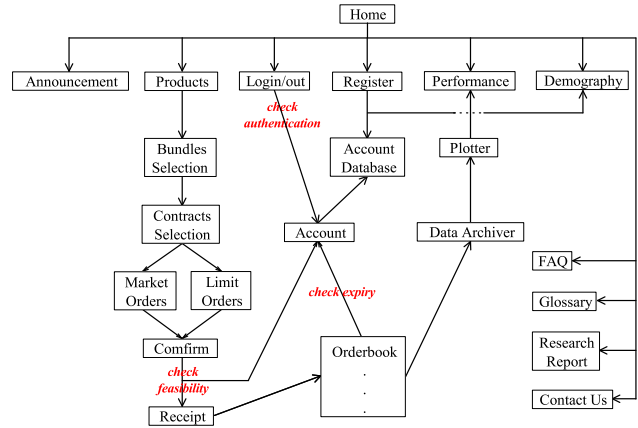}}
  \caption{Architecture of TAIPEX}
  \label{architecture}
\end{figure}

The concept of TAIPEX originates from the so-called Prediction Market\footnote{Prediction market
(also known as information market) is a market designed to run for the primary purpose of mining
and aggregating information scattered among traders.
The aggregation of the information will therefore be reflected in the form of market prices so as to make
predictions about specific future events.},
but there are dozens of ways to implement this idea.
We here design our platform TAIPEX as a 24-hour Web-based prediction market which facilitates the exchange of political
futures contracts whose liquidation prices are coupled to specific election outcomes~\cite{WI,QF,NT}.
Since October 2003, TAIPEX can be accessible at http://socioecono.phys.sinica.edu.tw.
Through the Internet,
anyone with a Web browser can participate in the market experiment by an on-line registration.
A private account with user-provided login name will be created for this registrant
and the platform will deposit an initial amount of virtual money into the account
immediately after successful registration.
The information about user demography, price fluctuation and accumulated volumes
are public to any browser irrespective of her registration or not,
however, only registered user with a valid account can trade in this market upon login.
In Fig.~\ref{architecture}, we depict the architecture of our platform.
The rules and instructions for the market experiment are detailed in the Announcement and FAQ sections.
Before starting to trade, participants would be asked to learn the rules and instructions.
Anyone who violates the rules to a certain extent, will be expelled from the experiment
and his account will be suspended.

In our design, a given futures contract is associated with the liquidation price which
equals to the percentage of votes that a candidate will get on the day of election.
As for the bundle, it consists of contracts for all candidates in the race
as well as for all the invalid ballots.
Afterward, the market prices of these contracts should sum up to around 100
if the traders behave rationally or if the market is efficient.
The minimal unit for the virtual money is set to one in our platform
and with the virtual money in the account, traders can buy bundles of futures contracts from
the platform for a guaranteed price or can buy/sell the contracts from/to the market directly.
One should keep in mind that
all the contracts in the account must be liquidated with the official results
from the election at the end of experiment,
therefore, the bundle price of 100 is fair since neither the user nor our platform loses.
Traders can place market or limit orders to either buy or sell futures contracts.
These bid (ask) orders are then stored in the order book maintained
by our platform which adopts the continuous double auction (CDA) as the
order matching and price discovery mechanism.
If no matches are found, market orders will expire immediately
while the limit orders will stay in the book and wait for further matches with new orders.
Nevertheless, these limit orders would either expire or be canceled by traders before the matches.
Once the matches are found, the transaction is made immediately,
followed by a corresponding balance within the traders' accounts.
The platform keeps all the details about the trading information
including the order submission, cancellation, expiration and transactions.

In addition to what we have mentioned above,
since the virtual money is used for the trading in our market,
we run the experiment as a tournament to encourage participants for further trading.
The traders can make investment decisions of their own free will to compete for
the monetary prizes provided by us,
however, only those whose ultimate wealth ranks in top ten will be rewarded.
The amount of monetary prizes varies for different experiments
but it is announced each time when a experiment begins.
Thus, the participant knows the payoff when her wealth finally ranks in the top ten.
At the end of experiment,
all the futures contracts are liquidated according to the official results of the election.
We will then announce the names of the winners on our platform
so that they can contact us to claim their rewards.
After identifying these winners,
we will transfer the money to the banking accounts they provide.
As for the source of participants, each time when we launch an experiment,
we first spread the news of this event throughout the internet on daily bases
and sometimes this news would also be reported by the press.
Therefore the number of traders and the market size will grow with time.
\subsection{Parameters and data set}
The data set analyzed in this study comes from a market experiment
targeting at 2008 Taiwan presidential election.
This experiment has lasted for 97 days, spanning from 2007/12/17 to 2008/03/22
which is the voting date for the election.
According to the official announcement from Central Election Commission of Taiwan,
we issued three futures contracts which consisted of
two candidates from two major political parties in Taiwan, namely Kuomintang (KMT)
and Democratic Progressive Party (DPP),
and one for any invalid ballots cast on the election day.
As we have mentioned earlier, a bundle consisting of three contracts
is provided at price 100 by our platform.
Each account began with no futures contracts
but with an amount of virtual money up to 10,000 units as the initial wealth.
During the experiment, the traders' accounts earned no interest
while no fees would be charged upon the transaction or order submission.
As for the rewards, the traders whose ultimate wealth ranking from top 1 to top 10
could get a monetary prize of 10,000, 8,000, 6,000, 5,000, 4,000,
2,500, 1,500, 1,000, 1,000 and 1,000 NT dollars\footnote{
The exchange rate during March 2008 is 1 USD $\approx$ 31 NTD.}, respectively.

\begin{figure}
  \resizebox{0.46\textwidth}{!}{\includegraphics{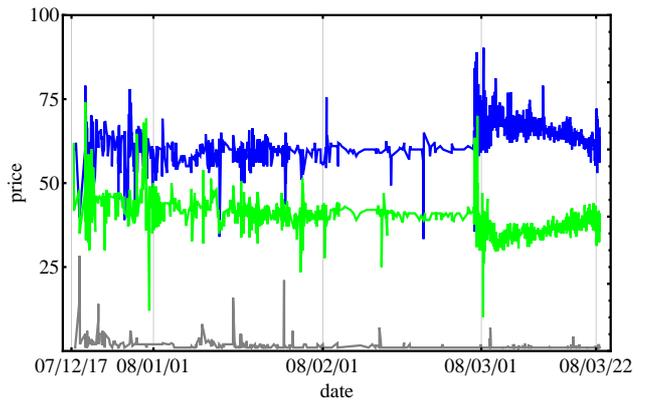}}
  \caption{The price time series during the whole experiment is shown.
  The contracts for candidates from KMT and DPP are drawn in blue and green, respectively,
  while the contract for invalid ballots is drawn in gray.}
  \label{price}
\end{figure}

The experiment has been monitored everyday to prevent from any possible incident
(system crash, blackout, hacker's attacks, etc.).
By the last day of experiment,
we accumulated 16,936 entries of transactions from 39,209 entries of orders submitted
by 2,095 effective traders and out of them,
there were only 1,985 traders who had made successful transactions.
According to the official result,
we liquidated each contract of KMT, DPP and invalid ballots at the price of 58, 41 and 1, respectively.
Fig.~\ref{price} shows the price time series covering the experiment.
The contracts for candidates from KMT and DPP are drawn in blue and green, respectively,
while the contract for invalid ballots is drawn in gray.
The intermittence of price spikes in the plot may originate in
a multiplicative process with additive noise which is supposed to yield
the power-law fluctuations~\cite{PT1,PT2}
or the so-called stylized facts in the financial market~\cite{EPBOOK}.
In an earlier work~\cite{QF},
we demonstrated that these stylized facts can be reproduced in our experiment.
Therefore, we assume that our market should act like a real financial market
and we further propose that our platform might be a good candidate
for performing the market experiment.

\section{Results and analysis}
In previous studies~\cite{NT,SF}, we demonstrated that the transaction networks
resulting from two parallel market experiments on our platform possess scale-free,
hierarchical and disassortative structure.
However, due to the low statistics in earlier experiments,
we can not make solid conclusions for neither the dynamics of network growth
nor the relation between individual wealth and network structure.
Nevertheless, in the latest experiment, we have higher statistics 
and finer details about the trading information.
We can therefore obtain the individual wealth dynamically or rebuild the transaction
networks in many different ways.
For example, the network in an arbitrary period of time or
the sub-network consisting of only the trading of specific contract can
be rebuilt without any difficulty.
\subsection{Topology of transaction networks}
\begin{figure}
\center{\resizebox{0.46\textwidth}{!}{\includegraphics{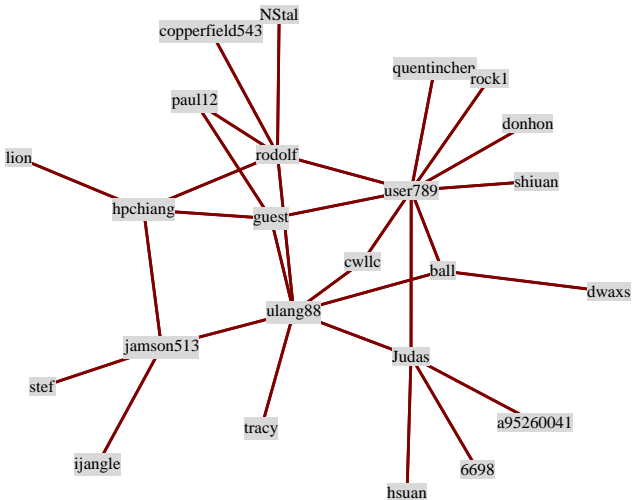}}}
  \caption{A snapshot of the transaction network in our experiment.
  This network consists of 24 nodes and 30 edges.}
  \label{networks}
\end{figure}

Each time a bid/ask order of trader $i$ is matched with an ask/bid order of trader $j$,
a transaction take place, following by an underlying exchange of virtual money and contracts
between the accounts of trader $i$ and $j$.
During the experiment, the platform keeps all the details of these kinds of exchange processes occurring in the market.
Therefore, from these records, one can readily reconstruct the so-called transaction networks
where the nodes represent the traders in our market and the edges between each pair of nodes
imply transactions among traders.
For demonstration, a typical appearance of the transaction network
during a given period of time is drawn in Fig.~\ref{networks}.
In this snapshot, we have a network consisting of 24 nodes and 30 edges.
The names on the nodes denote the user names for traders who have made at least one
transaction during this period while the edges depict the transactions among these traders.
From bunches of these snapshots, we notice that a hub (aggressive trader)
develops quickly in our networks.

\begin{figure}
\center{\resizebox{0.46\textwidth}{!}{\includegraphics{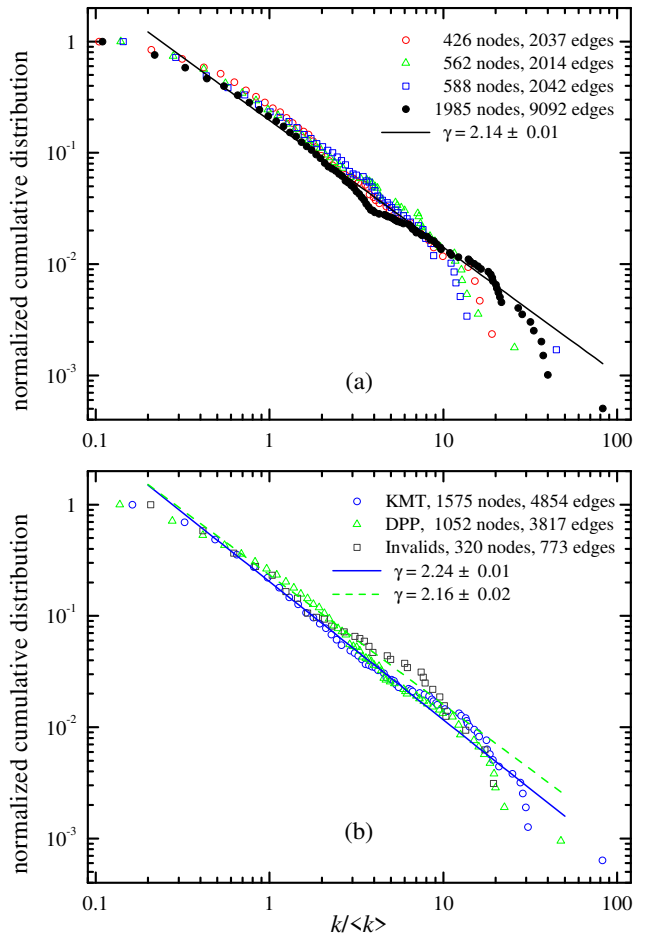}}}
  \caption{The normalized cumulative degree distribution
  ${\cal P}_{>}(k)$ of some selected transaction networks are plotted,
  where the solid and dash lines denote the power-law fit.
  The distributions for four different periods are shown in (a)
  while the results for three individual contracts (KMT, DPP and invalids) are plotted in (b).
  }
  \label{fig4}
\end{figure}

In the literature, a network is classified as a scale-free network provided that
its degree distribution function $P_d(k)$ decays as a power-law of the degree $k$.
Namely, we have $P_d(k)\sim k^{-\gamma}$ where the exponent $\gamma$ is a constant
ranging from 1 to 3 for different kinds of networks~\cite{CNBOOK}.
However, due to the low statics in the empirical data,
in alternative, one usually checks the cumulative degree distribution for the power-law behavior.
The cumulative degree distribution is defined as
\begin{equation}
  {\cal P}_{>}(k) \equiv \sum_{k'=k}^{k_{max}} P_d(k').
\end{equation}
and we have ${\cal P}_{>}(k) \sim k^{-(\gamma -1)}$ if $P_d(k)\sim k^{-\gamma}$.
In Fig.~\ref{fig4}, we plot ${\cal P}_{>}(k)$ of the transaction networks
reconstructed in two different ways,
namely, from the data over an arbitrary period of time
and from the data containing only the trading of a specific contract.
The largest network, shown as the black dot in Fig.~\ref{fig4}(a),
consisting of $1,985$ nodes and $9,092$ edges
is the one that spans across the whole time horizon (from 2007/12/17 to 2008/03/22).
The other three networks in Fig.~\ref{fig4}(a) with the size $n = 426, 562$ and $588$
are rebuilt from three non-overlapping periods,
while the results of another three sub-networks for individual contracts (KMT, DPP and invalids) are
shown in Fig.~\ref{fig4}(b).
One should notice that because these networks are of different sizes,
we have rescaled the degree $k$ to its average value $\langle k\rangle$ and
normalized ${\cal P}_{>}(k)$ in order to make the comparison among these distributions.
We observe that if we ignore the sudden drop of the distributions
near the end of large $k$, which might be due to the finite size effects,
all the distributions shown in Fig.~\ref{fig4} seem to collapse
onto a single distribution which can be well fitted by a power-law decay
with a corresponding exponent $\gamma$ ranging from $2.14$ to $2.24$.
Therefore, it is fair to assert that the transaction networks
in our market are scale-free with a similar exponent $\gamma$ when time evolves.
Even we extract a sub-network from the whole network,
the scale-free nature is still preserved.

\begin{figure}
\center{\resizebox{0.46\textwidth}{!}{\includegraphics{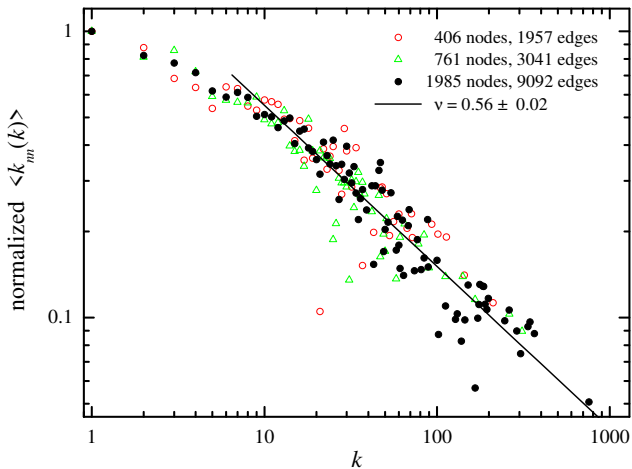}}}
  \caption{The normalized $\langle k_{nn}(k)\rangle$ 
  for transaction networks taken from three different periods of time.
  The solid line is a fit of the form $\langle k_{nn}(k)\rangle \sim k^{-\nu}$ with $\nu=0.52\pm 0.02$.
  }
  \label{fig5}
\end{figure}

In addition to the degree distribution,
we also work out the average nearest neighbors degree (ANND) to
see whether there exists the degree-degree correlation in our networks.
The ANND is defined as
\begin{equation}
  \langle k_{nn}(k)\rangle \equiv \sum_{k'=k}^{k_{max}} k' P_{d}(k'|k),
\end{equation}
where $P_{d}(k'|k)$ denotes the conditional probability
that an edge belonging to node with degree $k$ links to another one with degree $k'$~\cite{ANND}.
For uncorrelated networks, $\langle k_{nn}(k)\rangle = \langle k^2 \rangle/\langle k \rangle$,
independent of $k$.
In Fig.~\ref{fig5}, we plot $\langle k_{nn}(k)\rangle$
for the transaction networks in three different periods of time.
The $\langle k_{nn}(k)\rangle$ is shown to decay with increasing $k$,
therefore we conclude that our networks are always disassortative,
which means that the aggressive traders in our experiment tend not to
trade with each other.

\subsection{Correlation between degrees and wealth}
\begin{figure}
\center{\resizebox{0.46\textwidth}{!}{\includegraphics{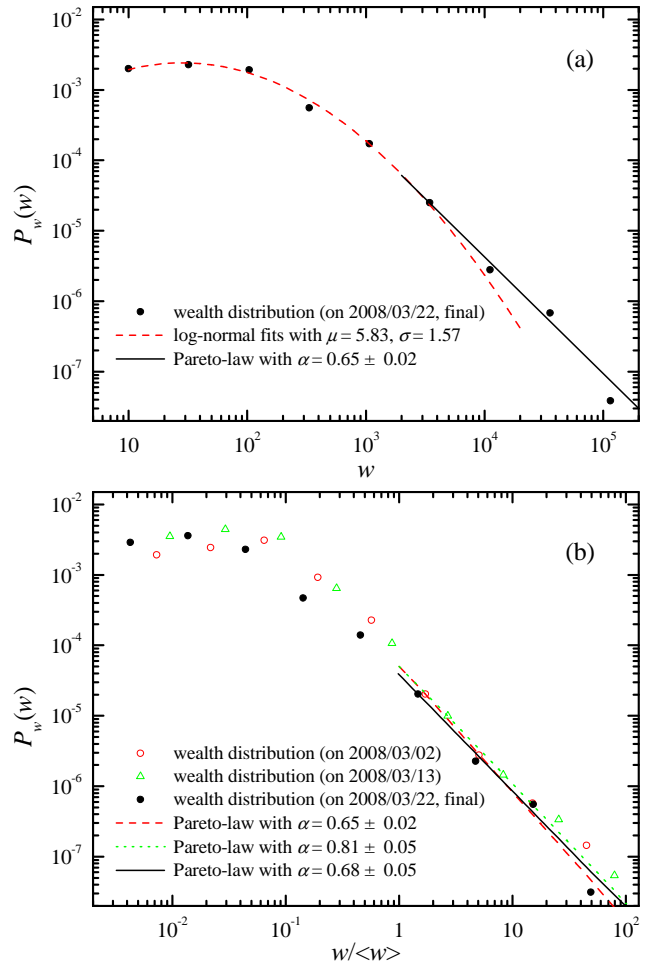}}}
  \caption{The wealth distribution at the end of experiment is plotted as the dots in (a),
  where the solid line denotes the Pareto-law tail with $\alpha=0.65\pm0.02$ while
  the dash line is the log-normal fit to the bulk.
  (b) shows the wealth distribution obtained on 2008/03/02 (circle), 2008/03/13 (triangle)
  and 2008/03/22 (dot) while the dash, dotted and solid lines denote the Pareto-law tails
  for the distribution on these dates.
  }
  \label{fig6}
\end{figure}

The wealth distribution functions $P_w(w)$ on 2008/03/22 (the end of experiment)
is shown in Fig.~\ref{fig6}(a),
where the richest guy earns 115,353 units of the virtual money in this experiment.
One can observe that the distribution for whom accumulate more than average wealth
($\langle w\rangle=$2,342 in this case)
follows a power-law decay with a Pareto index $\alpha=0.65\pm0.02$,
while the remaining bulk can be well fitted by a log-normal distribution.
To demonstrate the temporal evolution of the wealth distribution in our market,
we also plot, in Fig.~\ref{fig6}(b), the wealth distribution on three different dates.
The circle, triangle and dot denotes the distribution obtained on 2008/03/02, 2008/03/13 and
2008/03/22, respectively, while the wealth is rescaled to its average value $\langle w\rangle$
estimated on these dates.
One can observe that although the Pareto index $\alpha$ varies from $0.65$ to $0.81$
in Fig.~\ref{fig6}(b), the hybrid distribution (log-normal bulk and power-law tail)
of the wealth is still preserved.
We also attempt to figure out how $\alpha$ involves with time in our market,
but have not yet succeeded.

\begin{figure}
\center{\resizebox{0.46\textwidth}{!}{\includegraphics{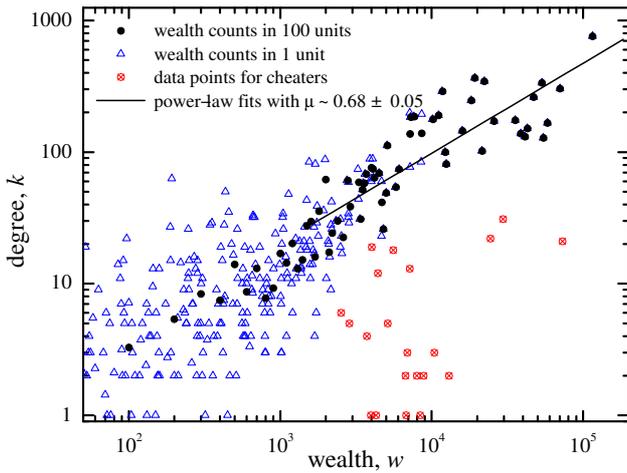}}}
  \caption{The relation between the degree and individual wealth is shown.
  The dot (triangle) denotes the wealth counted in units of 100 (1) while
  the circle with cross indicates the cheaters in our experiment.
  The solid line is the power-law fit (for $w$ beyond the average value $\langle w\rangle$)
  which follows the form $k \sim w^{\mu}$ with $\mu=0.68\pm 0.05$.
  }
  \label{fig7}
\end{figure}

To further study the relationship between the wealth accumulation and transaction networks,
we calculate the correlation between the wealth $w$ and degrees $k$.
The results are shown in Fig.~\ref{fig7} where
the dot (triangle) denotes the data for individual wealth
counted in units of 100 (1),
while the circle with cross indicates the cheaters\footnote{
The cheater here refers to whom benefits from the insider trading in our market.
The suspect cheaters are first distinguished from the large deviation to
the normal trend and then identified after checking their trading records.}
in our experiment.
It is shown that $w$ and $k$ are strongly correlated
above some critical values.
We can thus obtain a power-law fit for the data with
the individual wealth beyond the average wealth $\langle w\rangle$,
which reads as $k \sim w^{-\mu}$ with $\mu=0.68\pm 0.05$.
Summarizing all the findings so far, we have
\begin{eqnarray}
\label{eqn_dd}      P_{d}(k) &\sim& k^{-2.14\pm0.01}~~,\\
\label{eqn_wd}      P_{w}(w) &\sim& w^{-1.65\pm0.02} ~~~~\hbox{for $w>\langle w\rangle$},
\end{eqnarray}
and
\begin{eqnarray}
\label{eqn_kw}      k \sim w^{0.68\pm0.05} ~~~~~~\hbox{for $w>\langle w\rangle$}.
\end{eqnarray}
One may notice that, among the above equations,
the three fitting exponents $\gamma, \alpha$ and $\mu$ can be related to each other.
For instance, if we insert Eq.~\ref{eqn_kw} into Eq.~\ref{eqn_dd},
the value of $(0.68\pm0.05) \times (-2.14\pm0.01)$ falls between $-1.34$ and $-1.57$
which is roughly comparable to the exponent $-1.65\pm0.02$
for the wealth distribution in Eq.~\ref{eqn_wd}.
Although the relations are not exact,
we nevertheless shed the light on the possible explanation why Pareto's law persists
across economies.

\section{Conclusion}
From a market experiment with human agents who can make investment decisions of their own free will,
we explore the topology of transaction networks and demonstrate that
it is scale-free and disassortative for those with the size of the order of few hundred.
Even as time evolves, these nontrivial properties are still preserved.
We also observe that the wealth distribution in our market follows
the Pareto's law with a Pareto index $\alpha=0.65\pm0.02$ for the rich
and a log-normal distribution for the remaining bulk.
As the time evolves, the hybrid distribution of the wealth persists.
By further calculating the correlation between degrees and wealth,
we argue that in our experiment, the exponent $\alpha$ for individual wealth,
the exponent $\gamma$ for degree distribution and the exponent $\mu$ for wealth-degree correlation
can be roughly related to each other.
This finding may also be true for the cases in other real world markets.
Apart from that,
we have also noticed that by probing the wealth-degree correlation one can readily
identify the suspects of cheaters or the participants
who benefit from the insider trading in our market,
which might be useful against economic crimes in real financial markets.
With more and more data coming from the market experiments,
we believe that the interplay between the growth of transaction networks
and wealth accumulation process should be understood in the near future.
%
%

\end{document}